\documentclass[aps,pra,showpacs,groupedaddress,floatfix,showkeys]{revtex4}
\usepackage{graphicx}
\usepackage{amsmath,amssymb}
\usepackage{times,txfonts}
\usepackage{color}

\begin{document}

\title{Accuracy of the Faddeev Random Phase Approximation for Light Atoms}

\author{C. Barbieri}
\affiliation{Theoretical Nuclear Physics Laboratory, RIKEN Nishina Center,
2-1 Hirosawa, Wako, Saitama 351-0198 Japan}
\author{D. Van Neck}
\affiliation{Center for Molecular Modeling, Ghent University, 
Technologiepark 903, B-9052 Gent, Belgium}
\author{M. Degroote}
\affiliation{Center for Molecular Modeling, Ghent University, 
Technologiepark 903, B-9052 Gent, Belgium}

\date{\today}

\begin{abstract}

 The accuracy of the Faddeev random phase approximation (FRPA) method
is tested by calculating the total and ionization energies of a set
of light atoms up to Ar. 
 Comparisons are made with the results of coupled-cluster singles and
doubles~(CCSD), third-order algebraic diagrammatic construction [ADC(3)],
and with the experiment.
It is seen that even for two-electron systems, He and Be$^{2+}$,
the inclusion of RPA effects leads to satisfactory results and therefore
it does not over-correlate the ground state.
 The FRPA becomes progressively better for larger atomic numbers where
it gives $\approx$5~mH more correlation energy and it shifts
ionization potentials by 2-10~mH, with respect to its sister method ADC(3).
The corrections for ionization potentials consistently reduce
the discrepancies with the experiment.

\end{abstract}


\maketitle

%%%%%%%%%%%%%%%%%%%%%%%%%%%%%%%%%%%%%%%%%%%%%%%%%%%%%%%%%%%%%%%%%%%%%%%%%%%%%%%%%%%%%

\section{Introduction}
\label{intro}

Microscopic calculations of many-electron systems are plagued by the
exponential increase of degrees of freedom with increasing number of particles.
For this reason, {\em ab-initio} treatments are limited to mid-size
molecules with up to $\sim$100 electrons~\cite{Hel.04,Bartlett.07,Star.09}. 
On the other hand, the Kohn-Sham formulation~\cite{Koh.65} of density 
functional theory (DFT)~\cite{Hoh.64} incorporates
many-body correlations (in principle exactly),
while only single-particle equations must be solved.
 Due to this simplicity DFT is the only feasible approach in some modern 
applications of electronic structure theory. There is therefore a continuing 
interest in studying conceptual improvements and extensions to the DFT
framework~\cite{Bart.05,Mor.05,Pas.08,Gee.09}.
Existing DFT functionals can handle short-range inter-electronic correlations
quite well, while there is room for improvements in the description
of long-range effects~\cite{Yan.00,Fur.08}.
%
%Microscopic theory can help in improving the functionals, for example
%in {\em ab-inito} DFT~\cite{Bart.05,Mor.05} or in seeking expansions of
%the total energy in terms of the density matrix~\cite{Pas.08,Gee.09}.
%
An important issue is how to improve the treatment of long-range (van der Waals)
forces and dissociation processes, which could be approached
by combining DFT functionals with microscopic
calculations in the random phase approximation (RPA)~\cite{Dobson,Dio.04,Rom.09}.
% Several works have therefore focused on how to include microscopic RPA theory
%into DFT functionals~\cite{Dio.04,Rom.09}.
The RPA has also the interesting characteristic that it gives finite correlation
energies in metals and the uniform electron gas, where it appropriately screens
the Coulomb interaction at large distances~\cite{Mattuk,DicVan}.

In a recent publication we have considered the {\em ab-initio} calculation
of the Ne atom using Green's function theory in the so called Faddeev-RPA (FRPA)~\cite{Bar.07}.
This approach includes completely two-particle--one-hole (2p1h) and
two-hole--one-particle (2h1p) states in the self-energy.
Moreover, these configurations are grouped in terms of particle-vibration
couplings where the collective vibrations are calculated using RPA.
 The FRPA was originally proposed for studies of nuclear structure~\cite{DiB.04,Bar.06,Bar.04,Bar.09c,Bar.09d},
however, due to these characteristics it appears capable of treating
long-range correlations in both finite and extended electron systems.
The present work extends the calculations of Ref.~\cite{Bar.07} to other closed
(sub)shell systems and investigates the accuracy of FRPA for light atoms.

 Closely related to this study, is the development of the
quasi-particle (QP)-DFT formalism proposed in Ref.~\cite{Van.06}. 
In the QP-DFT the full spectral function is decomposed in the contribution
of the QP excitations, in the Landau-Migdal sense~\cite{Mig.67},
and a remainder or background part. 
Using a functional model for the energy-averaged background part,
it is possible to obtain a single-particle self-consistency problem that generates
the QP excitations.
 Such an approach is appealing since it contains the well-developed 
standard Kohn-Sham formulation of DFT as a special case, while  at 
the same time emphasis is put on the correct description of QPs. Hence, it can provide an improved 
description of the dynamics at the Fermi surface.  Given the close relation 
between QP-DFT and the Green's function formulation of many-body 
theory~\cite{FetWal,DicVan},  it is natural to employ {\em ab-initio} 
calculations in the latter formalism to investigate the structure 
of possible QP-DFT functionals. 
In doing so, one wishes to be able to calculate both atomic/molecular systems
and the homogeneous electron gas with the same approach.

For QPs associated with outer valence states in molecules,
the importance of a treatment that is consistent with at least
third-order perturbation theory was already pointed out by Cederbaum
and co-workers~\cite{Ced.77,Wal.81}.
 Such contributions were then included in the algebraic diagrammatic
construction method at third-order [ADC(3)]~\cite{Sch.82,Sch.83}.
The accuracy of the ADC approach has been tested in 
several studies for both ionization energies~(IEs)~\cite{Pern.04,Tro.05} and excited states~\cite{Tro.02,Star.09}.
 Based on perturbation theory arguments, ADC(3)
is expected to be comparable with coupled cluster singles and doubles (CCSD) for
correlation energies and  with coupled cluster singles doubles and triples (CCSDT)
for electron attachment and removal~\cite{Tro.05}.
The ADC(3) approach performs explicit configuration mixing between 2p1h
and 2h1p states which generate shake-up configurations of deeply bound orbits.
These states are mixed together by ADC(3) theory in a Tamm-Dancoff approximation (TDA) fashion.

Extended systems require a different approach since the TDA diverges
in the particle-hole (ph) channel due to the long-range part of the
Coulomb force.
In this cases, RPA becomes essential to obtain properly screened interaction
and one resorts to the GW method to include the coupling of 
electrons to ph-RPA phonons~\cite{Hed.65}.
This approach reproduces with high accuracy single-particle levels in
solids and atomic systems~\cite{Oni.02,Ary.98,Ver.06,Stan.09}
and self-consistent calculations yield the exact correlation
energy for the electron gas~\cite{Bar.EG1,Bar.EG2,Shi.96,Dew.05}. However,
the GW approximation is not free of troubles: while self-consistency
improves correlation energies it also deteriorates
the description of spectral quantities.
The incomplete treatment of Pauli correlations can lead to an
incorrect self-screening of the single-particle strength and
consequently to poor quasiparticle properties~\cite{Roman.09}.
 Deterioration of QP states is also found in atoms when seeking partial
improvements of Pauli exchange effects, as in the generalized GW (GGW)~\cite{Ver.06}.
This makes the GW approximation unsatisfactory for atomic and molecular systems.
Consistently with these results, Ref.~\cite{Bar.07} found that including
the sole ph-RPA---as done by the GW approximation---exaggerates correlations
for outer valence orbits.
To properly compute electron affinities (EAs) and IEs
one must also account for interactions in the particle-particle (pp)
and hole-hole (hh) channels.
When this was done through the FRPA, they corrected large part of the errors
of GW and reproduced the correct physics.

The FRPA formulation of Refs.~\cite{Bar.07,Bar.09c} includes the ADC(3) completely and
has similar computational requirements. At the same time, the explicit inclusion of RPA 
introduces the effects of ground state correlations in the phonons' spectra.
Thus it holds the promise for bridging calculations of finite and extended
electron systems.
Still, there remain a number of issues that need to be addressed before this
approach can be made useful for electronic structure calculations. 
First, it is not  {\em a-priori} guaranteed that this approach would behave
equally well for few-electron cases: while Pauli correlations are fully
included up to 2p1h/2h1p
level, the use of RPA still includes violations for more complex excitations.
Instabilities and self-screening problems induced by RPA may become more evident
in light systems.
Second, the inclusion of Pauli exchange at the 2p1h/2h1p introduces
technical complications with respect to the standard GW approach and
suitable computational schemes will need to be developed for applications
to extended systems.
This work addresses the first question by applying FRPA to light
closed sub-shell atoms.

The essential features of FRPA are reviewed in Sec.~\ref{formalism},
for the paper to be self-contained. References to the details
of the formalism are also given.
In our calculations we adopt gaussian basis sets and perform extrapolations
to the basis set limit. Hence, we precede the results with a discussion
of the accuracy in the extrapolations, in Sec.~\ref{results:conv}.
The results for total energies and IEs are given in Sec.~\ref{results:all}
and the major conclusions are summarized in Sec.~\ref{conclusions}.

\section{Formalism}
\label{formalism}

\begin{figure}
  \includegraphics[height=.25\textheight]{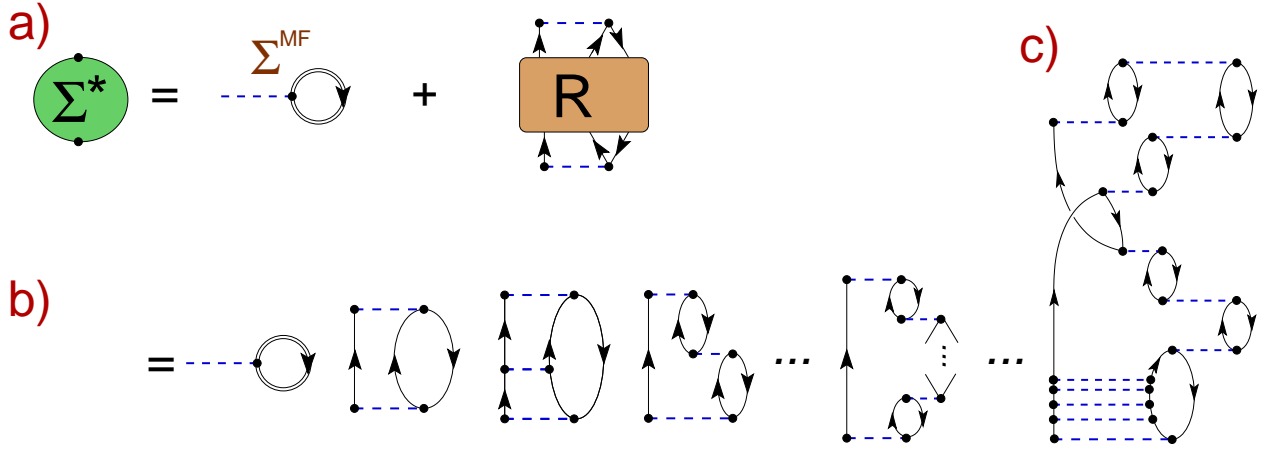}
%  \includegraphics[height=.21\textheight]{fig1ab}
%  \hspace{1.4cm}
%  \includegraphics[height=.19\textheight]{fig1c}
  \caption{The self-energy $\Sigma^\star(\omega)$ separates exactly into
    a mean-field term, $\Sigma^{MF}$, and the polarization propagator $R(\omega)$ for the
    2p1h/2h1p motion, as shown in a).  
    Dashed lines are antisymmetrized Coulomb matrix elements $V_{\alpha \beta, \gamma \delta}$,
   single lines represent the reference state (a HF propagator), and the double
   lines represent the correlated propagator of Eq.~(\ref{eq:g1}).
     Upon expansion of $R(\omega)$ in Feynman diagrams, one obtains the 
    series of diagrams b) for the self-energy.
     The diagram c) is another---more complicated---term appearing in the expansion
    of $R(\omega)$ that is also included in the FRPA contribution of Fig.~\ref{fig:faddex}.
    \label{fig:sigma_exp} }
\end{figure}

The theoretical framework of the present study is that of propagator theory.
The object of interest is the single-particle propagator~\cite{FetWal,DicVan},
\begin{equation}
 g_{\alpha \beta}(\omega) ~=~ 
 \sum_n  \frac{ 
          \langle {\Psi^N_0}     \vert c_\alpha        \vert {\Psi^{N+1}_n} \rangle
          \langle {\Psi^{N+1}_n} \vert c^{\dag}_\beta  \vert {\Psi^N_0} \rangle
              }{\omega - (E^{N+1}_n - E^N_0) + i \eta }  ~+~
 \sum_k \frac{
          \langle {\Psi^N_0}     \vert c^{\dag}_\beta  \vert {\Psi^{N-1}_k} \rangle
          \langle {\Psi^{N-1}_k} \vert c_\alpha        \vert {\Psi^N_0} \rangle
             }{\omega - (E^N_0 - E^{N-1}_k) - i \eta } \; ,
\label{eq:g1}
\end{equation}
where $\alpha$, $\beta$, ..., label a complete orthonormal basis set and
$c_\alpha$~($c^\dag_\beta$) are the corresponding second quantization 
destruction (creation) operators.
In these definitions, $\vert\Psi^{N+1}_n\rangle$, $\vert\Psi^{N-1}_k\rangle$ 
are the eigenstates, and $E^{N+1}_n$, $E^{N-1}_k$ the eigenenergies of the 
($N\pm1$)-electron system. Therefore, the poles of the propagator reflect the 
EAs and IEs.
%%electron affinities (EA) and ionization energies~(IE).
%%For a two-body hamiltonian,
Eq.~(\ref{eq:g1}) also yields the total binding energy via the
Migdal-Galitski\u{\i}-Koltun sum rule~\cite{bof.71,DicVan}.
The one-body Green's function is computed by solving the Dyson equation
(from hereafter, summations over repeated indices are implied)
\begin{equation}
 g_{\alpha \beta}(\omega) ~=~  g^{0}_{\alpha \beta}(\omega) ~+~
  %% \sum_{\gamma \delta} 
    g^{0}_{\alpha \gamma}(\omega)  \,
     \Sigma^\star_{\gamma \delta}(\omega) \,  g_{\delta \beta}(\omega) \; \; ,
\label{eq:Dys}
\end{equation}
where $g^0(\omega)$ is the propagator for a free particle (with only kinetic energy).
The irreducible self-energy $\Sigma^\star_{\gamma \delta}(\omega)$ acts
as an effective, energy-dependent, potential that can be written as
\begin{eqnarray}
  \Sigma^\star_{\alpha \beta}(\omega) &=&
 ~~~~~  \Sigma^{MF}_{\alpha \beta}   ~~~~~ ~+~ ~~~~~ \tilde\Sigma_{\alpha \beta}(\omega)
\label{eq:Sigma}
\\
 &=& \int \frac{d\omega}{2\pi i} \; V_{\alpha \gamma, \beta \delta} \,
     g_{\delta \gamma}(\omega) \, e^{-i\omega\eta^+}
  ~+~ \frac 1 4  V_{\alpha \lambda, \mu \nu} \,
        R_{\mu \nu \lambda, \mu' \nu' \lambda'}(\omega)\, V_{\mu' \nu', \beta \lambda'}\; ,
\nonumber
\end{eqnarray}
where $V_{\alpha \beta, \gamma \delta}$ are antisymmetrized Coulomb
matrix elements.
 In Eq.~(\ref{eq:Sigma}) we have emphasized the mean-field (MF) contribution
to the self-energy. This generalizes the Hartree-Fock (HF) potential by
replacing the Slater MF with the (correlated)
density matrix extracted from the dressed propagator~(\ref{eq:g1}).
The $\Sigma^{MF}$ is represented by the first diagram on the right hand
side in Figs.~\ref{fig:sigma_exp}a) and~\ref{fig:sigma_exp}b).
The remaining term, $\tilde\Sigma(\omega)$, accounts for deviations from
the mean-field and depends on the polarization propagator
$R(\omega)$ which involves the simultaneous propagation of 2p1h or 2h1p
{\em and higher} excitations.
Eq.~(\ref{eq:Sigma}) is represented in Fig.\ref{fig:sigma_exp}a) in terms
of Feynman diagrams.
 The polarization propagator $R(\omega)$ can also be expanded in terms
of Coulomb matrix elements and simpler propagators,
as shown in Figs.~\ref{fig:sigma_exp}b) and~\ref{fig:sigma_exp}c).
This approach also helps in identifying
key physics ingredients of the many-body dynamics. By truncating the expansion to a particular
subset of diagrams or many-body correlations, one can then construct suitable approximations to
the self-energy or seek for systematic improvements of the method.
 Moreover, since infinite sets of linked diagrams are summed
the approach is non-perturbative and satisfies the extensivity condition~\cite{Bartlett.07}.

\begin{figure}[t]
\includegraphics[width=.30\textheight,clip=true]{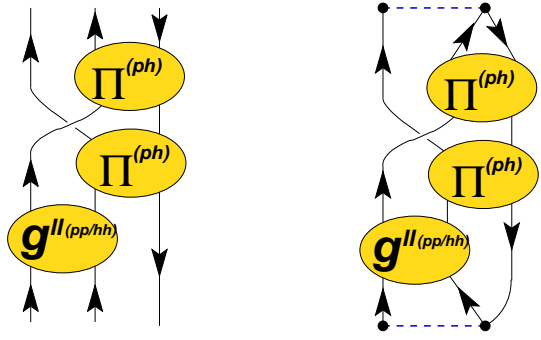}
\caption{ {\em Left:} Example of one of the diagrams
for $R(\omega)$ that are summed to all orders by means of the
FRPA Eqs.~(\ref{eq:FaddTDA}).
 Each of the ellipses represent an infinite sum of RPA
ladders~[$g^{II}(\omega)$] or rings~[$\Pi(\omega)$]---see Eqs.~(\ref{eq:Pi}),
(\ref{eq:g2}) and Fig.~\ref{fig:rpaeq}.
Contributions of all possible partial waves are included.
 {\em Right:} The corresponding contribution to the self-energy, obtained upon insertion
 of $R(\omega)$ into Eq.~(\ref{eq:Sigma}).}
\label{fig:faddex}
\end{figure}

 In the following we are interested in describing the coupling of
single-particle motion to ph, pp and hh collective
excitations of the system.
Following Ref.~\cite{Bar.01}, we first calculate the corresponding propagators
by solving the RPA equations in the ph and pp/hh
channels. These are then inserted in the self-energy by solving a set of
Faddeev equations to generate the 2p1h and 2h1p components of $R(\omega)$.
This approach is thus referred to as Faddeev RPA.
The whole procedure is equivalent to regrouping an infinite subset of Feynman
diagrams in the expansion of $R(\omega)$ as shown in Fig.~\ref{fig:faddex}.

The details of the FRPA approach are given in Refs.~\cite{Bar.01,Bar.07}
and are synthesized in Sec.~\ref{form_frpa} below.
For the discussion of the present work, it is sufficient to note
that including only ph propagators corresponds to the same physics
as the $GW$ approach~\cite{Hed.65}.
%%This is known to give accurate binding energies for the electron gas, where
%%the RPA is required to screen the long range Coulomb force.
The FRPA method goes beyond the $GW$ since it accounts completely for Pauli
correlations at the 2p1h/2h1p level and include the propagation of pp/hh
configurations. The latter interfere with ph phonons and have important
effects for ionization energies in finite systems~\cite{Bar.07}.
Contributions from all channels and in all possible partial waves are included
in FRPA, as it is required for a complete solution of the problem.
 For non-extended systems, it is sometimes possible to use the
TDA to calculate collective modes, instead of RPA.
It can be show that in this situation the FRPA reduces to the
third-order ADC of Schirmer~\cite{Sch.83}. In the following
we will refer equivalently to this approximation as either FDTA or ADC(3).
Throughout this paper, the dynamic self-energy $\tilde\Sigma(\omega)$
is calculated in terms of a HF reference state,
while $\Sigma^{MF}$ is derived self-consistently from
the dressed propagator, by iterating Eqs.~(\ref{eq:Dys}) and~(\ref{eq:Sigma}).

%%%%%%%%%%%%%%%%%%%%%%%%%%%%%%%%%%%%%%%%%%%%%%%%%%%%%%%%%%%%%%%%%%%%%%%%%
%%%%%%%%%%%%%%%%%%%%%%%%%%%%%%%%%%%%%%%%%%%%%%%%%%%%%%%%%%%%%%%%%%%%%%%%%

\subsection{The Faddeev random phase approximation method}
\label{form_frpa}

\begin{figure}
\includegraphics[width=0.40\textheight,clip=true]{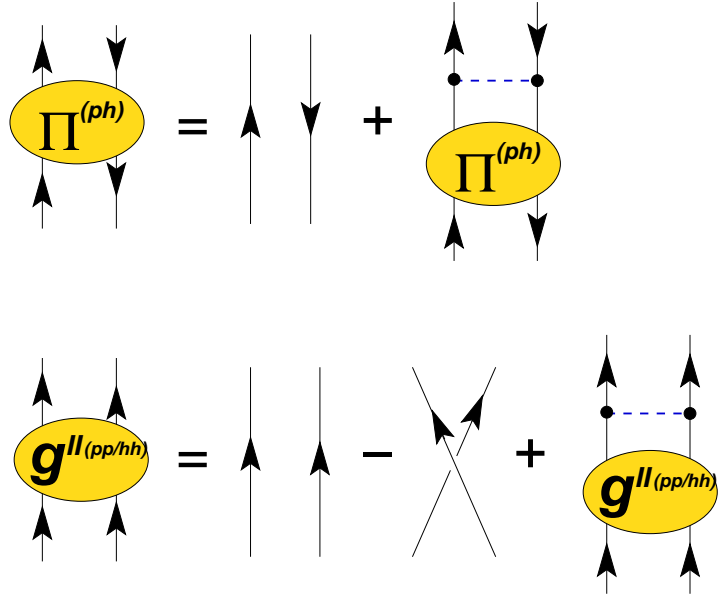}
\caption{ Diagrammatic equations for the polarization (above) 
and the two-particle (below) propagators. All time orderings are included
in order to generate the RPA series.}
\label{fig:rpaeq}
\end{figure}

Following Refs.~\cite{Bar.01,Bar.07}, we first consider the ph
polarization propagator that describes excited states of the $N$-electron system 
\begin{eqnarray}
 \Pi_{\alpha \beta , \gamma \delta}(\omega) &=& 
%% g_{\alpha \beta}(\omega) ~=~ 
 \sum_{n \ne 0}  \frac{  {\mbox{$\langle {\Psi^N_0} \vert $}}
            c^{\dag}_\beta c_\alpha {\mbox{$\vert {\Psi^N_n} \rangle$}} \;
             {\mbox{$\langle {\Psi^N_n} \vert $}}
            c^{\dag}_\gamma c_\delta {\mbox{$\vert {\Psi^N_0} \rangle$}} }
            {\omega - \left( E^N_n - E^N_0 \right) + i \eta } 
\nonumber \\
 &-& \sum_{n \ne 0} \frac{  {\mbox{$\langle {\Psi^N_0} \vert $}}
              c^{\dag}_\gamma c_\delta {\mbox{$\vert {\Psi^N_n} \rangle$}} \;
                 {\mbox{$\langle {\Psi^N_n} \vert $}}
             c^{\dag}_\beta c_\alpha {\mbox{$\vert {\Psi^N_0} \rangle$}} }
            {\omega + \left( E^N_n - E^N_0 \right) - i \eta } \; ,
\label{eq:Pi}
\end{eqnarray}
and the two-particle propagator that describes the addition/removal of two 
electrons
\begin{eqnarray}
 g^{II}_{\alpha \beta , \gamma \delta}(\omega) &=& 
%% g_{\alpha \beta}(\omega) &=& 
 \sum_n  \frac{  {\mbox{$\langle {\Psi^N_0} \vert $}}
                c_\beta c_\alpha {\mbox{$\vert {\Psi^{N+2}_n} \rangle$}} \;
                 {\mbox{$\langle {\Psi^{N+2}_n} \vert $}}
         c^{\dag}_\gamma c^{\dag}_\delta {\mbox{$\vert {\Psi^N_0} \rangle$}} }
            {\omega - \left( E^{N+2}_n - E^N_0 \right) + i \eta }
\nonumber \\  
&-& \sum_k  \frac{  {\mbox{$\langle {\Psi^N_0} \vert $}}
    c^{\dag}_\gamma c^{\dag}_\delta {\mbox{$\vert {\Psi^{N-2}_k} \rangle$}} \;
                 {\mbox{$\langle {\Psi^{N-2}_k} \vert $}}
                  c_\beta c_\alpha {\mbox{$\vert {\Psi^N_0} \rangle$}} }
            {\omega - \left( E^N_0 - E^{N-2}_k \right) - i \eta } \; .
\label{eq:g2}
\end{eqnarray}
%% We note that the expansion of $R(\omega)$ arises from applying the equations
%%of motion to the single-particle propagator~(\ref{eq:g1}), which is associated with the
%%ground state~$\vert\Psi^{A}_0\rangle$. Hence, all the Green's functions
%%appearing in this expansion will also be ground state based, including 
%%Eqs.~(\ref{eq:Pi}) and~(\ref{eq:g2}).
%% However, they contain in their Lehmann representations all the 
These Green's functions contain in their Lehmann representations all the 
relevant information regarding the excitation of ph and pp or hh modes.
In this work we are interested in studying the influence of collective
vibrations, which can be described within the RPA.
The propagators of Eqs.~(\ref{eq:Pi})
and~(\ref{eq:g2}) are then evaluated by solving the usual RPA equations,
which are depicted diagrammatically in Fig.~\ref{fig:rpaeq}.
In order to fulfill Pauli constraints in the expansion for $R(\omega)$
one must employ the generalized version of RPA, in which the Coulomb matrix elements
are the antisymmetrized ones.
 Since these equations reflect two-body correlations, they still have to be coupled
to an additional single-particle propagator, as in Fig.~\ref{fig:faddex}, to obtain
the corresponding approximation for the 2p1h and 2h1p components
of $R(\omega)$. This is achieved by solving two separate
sets of Faddeev equations, as discussed in Ref.~\cite{Bar.01}.
%%
%The Faddeev equations  also ensure that the Pauli principle is correctly
%taken into account at the  2p1h and 2h1p level. 

%%%%%%%%%%%%%%%%%%%%%%%%%%%%%%%%%%%%%%%%%%%%%%%%%%%%%%%%%%%%%%%%%%%%%%%%%%
%%%%%%%%%%%%%%%%%%%%%%%%%%%%%%%%%%%%%%%%%%%%%%%%%%%%%%%%%%%%%%%%%%%%%%%%%

 The $R(\omega)$ contains two separate terms that propagate either
2p1h or 2h1p states.
 Taking the 2p1h case as an example, one can further split $R^{(2p1h)}(\omega)$
again in three different components $\bar{R}^{(i)}(\omega)$ ($i=1,2,3$)
that differ from  each other by the last pair of lines that interact in
their diagrammatic expansion,
\begin{equation}
%\begin{eqnarray}
    \bar{R}^{(2p1h)}_{\alpha \beta \gamma , \mu \nu \lambda}(\omega) =
 \left[ {G^0}^>_{\alpha \beta \gamma , \mu \nu \lambda}(\omega)
        - {G^0}^>_{\beta \alpha \gamma ,  \mu \nu\lambda}(\omega) \right]
  + \sum_{i=1,2,3}     \bar{R}^{(i)}_{\alpha \beta \gamma , \mu \nu \lambda}(\omega)
  \; ,
\label{eq:faddfullR}
%\end{eqnarray}
\end{equation}
where ${G^0}^>(\omega)$ is the 2p1h propagator for three non-interacting lines.
These components are solutions of the following set of
Faddeev equations~\cite{Fad.61}
\begin{eqnarray}
  \lefteqn{
  \bar{R}^{(i)}_{\alpha  \beta  \gamma  ,
           \mu     \nu    \lambda   }(\omega) 
    ~=~ {G^0}^>_{\alpha  \beta  \gamma  ,
                 \mu'    \nu'   \lambda' }(\omega) ~
   \Gamma^{(i)}_{\mu'   \nu'     \lambda'  ,
                 \mu''  \nu''    \lambda'' }(\omega) }
        \hspace{.1in} & &
\nonumber  \\
  & \times & ~
  \left[ \bar{R}^{(j)}_{\mu''   \nu''  \lambda''  ,
                  \mu     \nu    \lambda    }(\omega) ~+~
         \bar{R}^{(k)}_{\mu''   \nu''  \lambda''  ,
                  \mu  \nu    \lambda    }(\omega)
  \right.
\label{eq:FaddTDA}  \\
  & & ~ ~+~ \left.
      {G^0}^>_{\mu''    \nu''   \lambda''   ,
               \mu      \nu     \lambda   }(\omega)
    - {G^0}^>_{\nu''    \mu''   \lambda''   ,
               \mu      \nu     \lambda   }(\omega)
       \right]  \; , ~ ~ ~ ~ i=1,2,3 ~ ,
%       \right]  \;  ,
\nonumber
\end{eqnarray}
where ($i,j,k$) are cyclic permutations of ($1,2,3$).
The interaction vertices $\Gamma^{(i)}(\omega)$  contain the couplings of 
a ph, see Eq.~(\ref{eq:Pi}), or pp/hh, see Eq.~(\ref{eq:g2}), excitation
to a freely propagating line. 
The propagator $R(\omega)$ which we employ in Eq.~(\ref{eq:Sigma}) is finally obtained by 
\begin{equation}
  R^{(2p1h)}_{\alpha  \beta  \gamma  ,
              \mu     \nu    \lambda   }(\omega)  ~=~
  U_{\alpha  \beta  \gamma , \mu'    \nu'   \lambda' } \;
   \bar{R}^{(2p1h)}_{\mu'   \nu'     \lambda'  ,
             \mu''  \nu''    \lambda''}(\omega)
    \;  U^\dag_{\mu''  \nu''    \lambda'' , \mu     \nu    \lambda} \; ,
\label{eq:URU}
\end{equation}
where the correction vertex $U$ ensures consistency with perturbation theory
up to third order.
  The explicit formulae for $\Gamma^{(i)}(\omega)$ and~$U$ are given in terms of
the propagators of Eqs.~(\ref{eq:Pi}), (\ref{eq:g2}) and the
interaction~$V_{\alpha \beta , \gamma \delta}$. They are discussed in detail
in  Ref.~\cite{Bar.07}.
The calculation of the  2h1p component of $R(\omega)$ follows completely analogous steps.
 
The present formalism includes the 
effects of  ph and  pp/hh motion simultaneously, 
while allowing interferences between these modes. 
These excitations
 are evaluated here at the RPA level and are then coupled to each other by 
solving Eqs.~(\ref{eq:FaddTDA}). This generates an infinite resummation of diagrams,
including the one displayed in Fig.~\ref{fig:faddex}.
When one uses TDA to approximate phonons this expansion reduces
to standard configuration mixing between 2p1h or 2h1p states.
As already noted, with the present formulation of
Eqs.~(\ref{eq:faddfullR}-\ref{eq:URU}) the FTDA turns out to be exactly
equivalent to the ADC(3) formalism of Ref.~\cite{Sch.83}. 
%
%%The FRPA scheme includes additional effects due to ground state correlations
%%that become important for extended systems.
%%%
%%In addition, one can in principle employ dressed single-particle propagators
%%in these equations to generate a fully self-consistent solution, 
%%as done in Refs.~\cite{Bar.02,Bar.06} for valence orbits around~$^{16}$O.

\section{Results}
\label{results}

We considered a set of neutral atoms and ions corresponding to 
closed shell and subshell configurations with Z$\leq$18. 
The calculations of the smallest systems (He, Be and Be$^{2+}$) were performed
using the correlation-consistent polarization valence gaussian bases, cc-pvXz,
of increasing quality from double- to quintuple-zeta (X=2-5).
For the larger atoms it was found that a sizable fraction 
of the correlation energy is lost with similar bases. The remaining systems were therefore
calculated with the corresponding core-valence bases, cc-pcvXz, which include additional
compact gaussians to improve the description of the core electrons.
This choice was seen to speed up the convergence and led to accurate results
for these atoms~\footnote{The augmented version of the bases (aug-cc-pvXz)
were also tested and gave no sizable improvement for the quantities
being considered in this work.}.
The correction to the correlation energies induced by the extra core orbits
increases with the number of electrons and it was found to be $\approx$40~mH for Ne
and $\approx$300~mH for Mg.

The bases for the Be$^{2+}$~(Mg$^{2+}$) ions were obtained from
the cc-p(c)vXz sets for He~(Ne) but scaling the corresponding
single-particle orbits to correct for the different atomic number,
\begin{equation}
 \phi^i_{{\rm Be}^{2+}/{\rm Mg}^{2+}}\big(r\big)
     ~~ \propto ~~ \phi^i_{He/Ne}\big(r\frac{Z}{N}\big) 
\label{scale_ions}
\end{equation}
where Z is the nuclear charge and N=Z-2 the number of electrons.

 Correlation and ionization energies were computed with both the FTDAc/ADC(3)c
and the FRPAc methods. In this notation, the letter 'c' indicates the self-consistent
treatment of the sole MF diagram in the self-energy [first diagram on the
r.h.s. in  Figs.~\ref{fig:sigma_exp}a) and~\ref{fig:sigma_exp}b)].
In other words, $\Sigma^{MF}$ is renormalized by evaluating it directly in terms
of the fully correlated propagator, instead of the reference state. 
This aspect is important since it consistently includes all the PT contributions up
to third-order and more.
 The ground states energies were also compared to the results of CCSD. In all cases the
HF wave functions (calculated for each basis set) were used as the reference state.

\subsection{Convergence}
\label{results:conv}

Total binding energies predicted by both Green's function theory and 
CCSD are shown with the results of full configuration interaction (FCI)
in Tab.~\ref{tab:1}.
For Ne atom in cc-pcvDz, Green's functions and CCSD agree with
each other and deviate from the exact result by less than 2~mH.
FRPA gives just a very small correction but it halves the discrepancy
between FTDA and FCI. The total correlation energy for this basis is 233~mH.
The atom of Be is the most difficult case among those discussed here
due to the fact that this is not a good closed-shell system.
 In this system, a near degeneracy between the 2s and 2p orbitals leads
to very soft excitations of the J$^\pi$=1$^-$, S=1 states and drives the
ph-RPA equations close to instability.
To avoid this, the FRPA was solved by employing the TDA approximation
of the polarization propagator, Eq.(\ref{eq:Pi}), in this channel alone
(all other partial waves were treated properly in RPA).
The resulting correlation energies agree with FTDAc/ADC(3)c, showing
that RPA is not crucial for this small system neither it introduces
spuriousities by over-correlating the ground state.
FCI calculations of Be were possible for all bases up to quintuple-zeta and 
are reproduced by CCSD with high accuracy.
However, FTDAc/ADC(3)c and FRPAc are consistently behind by about 9~mH,
corresponding to 10\% of the total correlation energy.
This is the most serious discrepancy we obtain in this work and suggest
a limitation of the FRPA---in its present form---for near-degenerate systems.
To overcome this, it may be necessary to introduce self-consistency
in the polarization propagator $R(\omega)$, to account for orbit relaxation,
or to improve the treatment of the excitation spectrum of the
polarization propagator beyond bare ph states~\cite{Bar.03}~%
\footnote{In the ADC language, this means adding fifth-order terms that are
introduced at the ADC(5) level.}. 
The close agreement between FTDAc/ADC(3)c and FRPAc in Tab.~\ref{tab:1}
is a welcome feature since for a few electrons in the Be atom one should
not expect collectivity effects to be important.
Although the RPA approximation is not meant for few-body systems, this 
result (and the one for He, below) shows that it can be safely applied
also in this regime without serious consequences.
The usual issues of RPA for cases of near degeneracy remain and may lead
to instabilities in certain channels, as just described.
%This can happen for systems of any size.

 Extrapolations to the basis set limit were obtained from two consecutive 
sets according to
\begin{equation}
E_X = E_\infty + A X^{-3} \; , 
\label{extrap}
\end{equation}
where $X$ is the cardinal number of the basis. This relation is known to 
give proper extrapolations for correlation energies~\cite{Hel.04}.
 In Sec.~\ref{results:all}, we will apply it to ionization energies as well,
remembering that these are also differences between eigenenergies.
Table~\ref{tab:2} gives some examples of the calculated binding energies
for all bases sizes  and  shows the convergence of the extrapolated results.
In the smallest systems, up to Ne, we find changes of less than 2~mH between
the last two extrapolations ($X$=$T$,$Q$ and $X$=$Q$,$5$). This number can be
taken as a measure of the uncertainty in reaching the basis set limit.
For the larger atoms Mg is the one that converges more slowly, with a difference
of 10~mH (we found 7~mH for Ar). Calculations with X=6 are beyond present computational 
capabilities. However, given the fast convergence with increasing cardinal number, it
appears safe to assume an uncertainty of $\leq$5~mH for these cases.

%%%%%%%%%%%%%%%%%%%%%%%%%%%%%%%%%%%%%%%%%%%%%%%%%%%%%%%%%%%%%%%%%%%%%%%%%
\begin{table*}[h]
\begin{ruledtabular}
\begin{tabular}{lccccc}
$E_{\mbox{tot}}$      &    Ne        &   \multicolumn{4}{c}{Be} \\
                      & ------------ &   \multicolumn{4}{c}{---------------------------------------------}\\
                      &   cc-pcvDz   &  cc-pvDz  & cc-pvTz  & cc-pvQz  & cc-pv5z \\
\hline
ADC(3)c/FTDAc         &   -128.7191  &    -14.6089  &	-14.6154  &   -14.6314  &   -14.6375 \\
FRPAc                 &   -128.7210  &    -14.6084  &	-14.6150  &   -14.6310  &   -14.6371 \\
CCSD                  &   -128.7211  &    -14.6174  &	-14.6236  &   -14.6396  &   -14.6457 \\
\\
full CI               &   -128.7225  &    -14.6174  &	-14.6238 &    -14.6401  &   -14.6463\\
\end{tabular}
\end{ruledtabular}
%\end{center}
\caption[]{Total binging energies (in Hartrees) for Ne and Be obtained for cc-p(c)vXz bases
of different sizes. The results obtained with ADC(3)c and FRPAc (with partial self-consistency
in the MF diagram) and with the CCSD methods are compared to FCI calculations.} 
\label{tab:1}
\end{table*}
%%%%%%%%%%%%%%%%%%%%%%%%%%%%%%%%%%%%%%%%%%%%%%%%%%%%%%%%%%%%%%%%%%%%%%%%%

%%%%%%%%%%%%%%%%%%%%%%%%%%%%%%%%%%%%%%%%%%%%%%%%%%%%%%%%%%%%%%%%%%%%%%%%%
\begin{table*}[h]
\begin{ruledtabular}
\begin{tabular}{rlccccr}
\multicolumn{2}{c}{$E_{\mbox{tot}}$}
                      & cc-p(c)vDz  & cc-p(c)vTz  & cc-p(c)vQz  & cc-p(c)v5z  & Experiment \\
\hline
Be:& calc.            &   -14.6084  &   -14.6150  &   -14.6310  &   -14.6371  &    -14.6674 \\
   & extrap.          &             &   -14.6178  &   -14.6427  &   -14.6436  &     \\
\\
Ne:& calc.            &   -128.7210 &   -128.8643 &    -128.9079	 &   -128.9226      &  -128.9383   \\
   & extrap.          &             &   -128.9246 &    -128.9397     &   -128.9381   \\
\\
Mg:& calc.            &   -199.8147 &   -199.9507 &    -200.0033     &	-200.0271      &   -200.054  \\
   & extrap.          &             &   -200.0080 &    -200.0417     &	-200.0519   \\
\end{tabular}
\end{ruledtabular}
%\end{center}
\caption[]{Convergence of binding energies (in Hartrees) in the FRPAc approach.
First lines: total energies calculated in  using double (X=D) to quintuple (X=5)
valence orbits basis sets. Second lines: results extrapolated from two consecutive
sets using Eq.~(\ref{extrap}).
 The Be atom was calculated with the cc-pvXz bases, while Ne and Mg were done
using cc-pcvXz.
 The experimental energies are from Refs.~\cite{Cha.96,WebBE}.} 
\label{tab:2}
\end{table*}
%%%%%%%%%%%%%%%%%%%%%%%%%%%%%%%%%%%%%%%%%%%%%%%%%%%%%%%%%%%%%%%%%%%%%%%%%

\subsection{Ground states and ionization energies of simple atoms}
\label{results:all}

Table~\ref{tab:3} shows the ground state energies extrapolated
from $X$=$Q$,$5$ for both Green's function and CCSD methods.
These are compared to the corresponding Hartree-Fock results and
the experiment. The empirical values are from
Refs.~\cite{Dav.91,Cha.96,WebBE} and have been corrected by
subtracting relativistic effects.
 The CCSD results for He and Be$^{2+}$ are equivalent to FCI, from which we
see that FRPAc misses 1~mH, or 2\%, of the correlation energy of He.
In larger systems FRPAc explains at least 99\% of the correlation energies
and all calculations, including CCSD, agree with the experiment
within the uncertainty expected from basis extrapolation. 
For Z$\geq$10, the inclusion of RPA phonons predicts about 5~mH more 
binding than the corresponding FTDAc/ADC(3)c.
 The atom of Be is the only exception to this trend as already noted above.
In this case the 9~mH difference between FRPAc and CCSD is seen also in the 
basis limit. 
 Based on the agreement between FCI and CCSD in Tab.\ref{tab:1}, the remaining
discrepancy with the experiment ($\approx$15~mH) may be due the basis set employed
which is probably not capable to accommodate the relevant correlation effects.
 We have attempted FRPAc calculations with the aug-cc-pvXz bases which should 
allow for a better description of the valence orbits but without any appreciable
change in the results.

The Ne atom was also computed in the FRPA approach by using a Hartree-Fock basis
with a discretized continuum~\cite{Bar.07}. The size of the basis set was chosen
as large as possible to approach the basis set limit, however, the set was 
optimized for the description of IEs and EAs rather than
for treating core orbits.
 The total binding energy obtained is 128.888~H, away from the basis set limit of
Tab.~\ref{tab:3} and the experiment.

%%%%%%%%%%%%%%%%%%%%%%%%%%%%%%%%%%%%%%%%%%%%%%%%%%%%%%%%%%%%%%%%%%%%%%%%%
\begin{table*}[h]
\begin{ruledtabular}
\begin{tabular}{lcccccccccc}
  &  & Hartree-Fock
  &  & FTDAc
  &  & FRPAc
  &  & CCSD
  &  & Experiment  \\
\hline
He         & &    -2.8617 (+42.0)  & &  -2.9028 (+0.9)   & &   -2.9029 (+0.8)  & &   -2.9039  (-0.2) & &   -2.9037 \\
Be$^{2+}$  & &   -13.6117 (+43.9)  & &  -13.6559 (-0.3)  & &  -13.6559 (-0.3)  & &  -13.6561  (-0.5) & &  -13.6556 \\
Be         & &   -14.5731 (+94.3)  & &  -14.6438 (+23.6) & &  -14.6436 (+23.8) & &  -14.6522 (+15.2) & &  -14.6674 \\
Ne         & &  -128.5505 (+387.8) & &  -128.9343 (+4.0) & & -128.9381 (+0.2)  & & -128.9353  (+3.0) & & -128.9383 \\
Mg$^{2+}$  & &  -198.83 7 (+444)   & &  -199.226  (-5)   & & -199.228  (-7)    & & -199.225 (-4)     & & -199.221 \\
Mg         & &  -199.616  (+438)   & &  -200.048  (+6)   & & -200.052  (+2)    & & -200.050 (+4)     & & -200.054 \\
Ar         & &  -526.820  (+724)   & &  -527.543  (+1)   & & -527.548  (-4)    & & -527.536 (+8)     & & -527.544 \\
\\
$\sigma_{rms}$ [mH] & &    392     & &     9.5 (3.6)     & &    9.5 (3.4)      & &    6.9 (4.2)      & & \\
\end{tabular}
\end{ruledtabular}
\caption{Hartree-Fock, ADC(3)c/FTDAc, FRPAc and CCSD binding energies (in Hartrees)
extrapolated from the cc-p(c)vQz and cc-p(c)v5z basis sets.
 He, Be$^{2+}$ and Be where calculated with the cc-pvXz bases, while cc-pcvXz bases were used
for the remaining atoms.
 The deviations from the experiment are indicated in parentheses (in mHartrees).
 The experimental energies are from Refs.~\cite{Dav.91,Cha.96,WebBE}.
 The {\em rms} errors in parentheses are calculated by neglecting the Be results.
} 
\label{tab:3}
\end{table*}
%%%%%%%%%%%%%%%%%%%%%%%%%%%%%%%%%%%%%%%%%%%%%%%%%%%%%%%%%%%%%%%%%%%%%%%%%

Ionization energies are shown in Tab.~\ref{tab:4}, together with the predictions
from Hartree-Fock theory and the second-order
self-energy (obtained by retaining only the first two diagrams of Fig.~\ref{fig:sigma_exp}b).
Second-order corrections account for a large part of correlations but still
lead to sizable errors. The additional correlations included in the present calculations
appear to reduce this error substantially. 
The FTDAc/ADC(3)c results give a measure of the
importance of a treatment that is consistent with at least third order
perturbation theory~\cite{Wal.81}.
 Corrections are particularly large for states with higher ionization
energies where the density of 2h1p states is increased. Since configuration
mixing among these states is not introduced by strict second-order
perturbation theory, calculations at least at the level of FTDAc are
required in these cases. Configuration mixing among the 2h1p states reduces
the errors in
%the $2p$ quasiparticle energy of Ar by a factor of two and
the $1s$ state in Be by a factor of five.
Another effect is the fragmentation of the $3s$ orbit of Ar. Second-order
calculations predict this as a quasiparticle state 36~mH away from
the empirical energy that carries 0.81 of the total orbit's intensity. A small
satellite state with relative intensity of 0.10 is calculated at larger separation energies.  
The mixing with 2h1p configurations corrects the energies of both
peaks and redistributes their strengths more correctly.
For the FRPAc calculation the peak at 1.065~H has
intensity of 0.61, close to the experimental value (0.55 found at 1.075~H~\cite{McC.89}).
 The second peak is obtained at 1.544~H and carries the remaining strength of the original
quasiparticle.

Adding the effects of collective RPA phonons has a larger impact on ionization
than on correlation energies. Almost all the IE calculated shift closer to the 
experimental values by a few~mH.  The only exceptions are the two-electron He atom, where
the RPA approach may tend to overestimate correlations, and the first ionization of 
Be, where soft excitations invalidate the RPA.
In general, the $rms$ error for the valence orbits of Tab.~\ref{tab:4}
%---with 6~H ionization energy or less---
lowers from 13.7 to 10.6~mH, passing from FTDAc to FRPAc.

The FRPAc IE of the Ne atom were also computed in the discretized continuum basis
of Ref.~\cite{Bar.07}.  The first and second ionization energies obtained are
0.801 and 1.795~H, in good agreement with the extrapolations of Tab.~\ref{tab:4}.
This gives us confidence on applying Eq.~(\ref{extrap}) also for quasiparticle states.

%%%%%%%%%%%%%%%%%%%%%%%%%%%%%%%%%%%%%%%%%%%%%%%%%%%%%%%%%%%%%%%%%%%%%%%%%
\begin{table*}[h]
\begin{ruledtabular}
\begin{tabular}{llcccccccccc}
& & & Hartree-Fock
  & & (2$^{nd}$  order)c
  & & FTDAc/
  & & FRPAc
  & & Experiment~\cite{NIST,Tho.01}   \\
& & &  & &  & & ADC(3)c  & &   & &   \\
\hline
He:        & 1s &  &  0.918 (+14)   &  &  0.9012 (-2.5) &  &  0.9025 (-1.2)  &  &  0.9008 (-2.9)   &  &  0.9037 \\
\\
Be$^{2+}$: & 1s &  &  5.6672 (+116) &  &  5.6542 (-1.4) &  &  5.6554 (-0.2)  &  &  5.6551 (-0.5)   &  &  5.6556 \\
\\
Be:        & 2s &  &  0.3093 (-34)  &  &  0.3187 (-23.9) & &  0.3237 (-18.9) &  &  0.3224 (-20.2) &  &  0.3426 \\
           & 1s &  &  4.733 (+200)  &  &  4.5892 (+56)  &  &  4.5439 (+11)   &  &  4.5405 (+8)    &  &  4.533 \\
\\
Ne:        & 2p &  &  0.852 (+57)   &  &  0.752 (-41)   &  &  0.8101 (+17)   &  &  0.8037 (+11)   &  &  0.793 \\
           & 1s &  &  1.931 (+149)  &  &  1.750 (-39)   &  &  1.8057 (+24)   &  &  1.7967 (+15)   &  &  1.782 \\
\\
Mg$^{2+}$: & 2p &  &  3.0068 (+56.9)&  &  2.9217 (-28.2) & &  2.9572 (+7.3)  &  &  2.9537 (+3.8)  &  &  2.9499 \\
           & 1s &  &  4.4827        &  &  4.3283        &  &  4.3632         &  &  4.3589         &  &   \\
\\
Mg:        & 3s &  &  0.253 (-28)   &  &  0.267 (-14) &  &  0.272 (-9)     &  &  0.280 (-1)     &  &  0.281 \\
           & 2p &  &  2.282 (+162)  &  &  2.117 ( -3) &  &  2.141 (+21)    &  &  2.137 (+17)    &  &  2.12  \\
\\
Ar:        & 3p &  &  0.591 (+12)   &  &  0.563 (-16)  &  &  0.581 (+2)     &  &  0.579 ($\approx$0)&& 0.579 \\
           & 3s &  &  1.277 (+202)  &  &  1.111 (+36)  &  &  1.087 (+12)    &  &  1.065 (-10)    &  &  1.075 \\
           & 3s &  &                &  &  1.840        &  &  1.578          &  &  1.544          &  &     \\
%%         & 2p &  &  9.571 (+411)  &  &  9.130 (-30)  &  &  9.235 (+75)    &  &  9.219 (+59)    &  &  9.160 \\
\\
$\sigma_{rms}$ [mH] & & &  81.4  &  & 29.3 &  &     13.7        &  &     10.6    &  & \\
%%$\sigma_{rms,~IE\,<\,6 H}$ [mH] & & &  81.4  &  & 29.3 &  &     13.7        &  &     10.6    &  & \\
\end{tabular}
\end{ruledtabular}
  \caption{Ionization energies obtained with Hartree-Fock, second-order perturbation
theory for the self-energy, FTDAc and with the full Faddeev-RPAc (in Hartrees).
 All results are extrapolated from the cc-p(c)vQz and cc-p(c)v5z basis sets (see Table 
\ref{tab:3}). The deviations from the experiment (indicated in parentheses) and the 
{\em rms} errors are given in mHartrees.
 The experimental energies are from Ref.~\cite{NIST, Tho.01}.
 %%The {\em rms} are given for the outer valence orbitals with IEs $<$6~H (excluding the 2p of Ar).
 } 
\label{tab:4}
\end{table*}
%%%%%%%%%%%%%%%%%%%%%%%%%%%%%%%%%%%%%%%%%%%%%%%%%%%%%%%%%%%%%%%%%%%%%%%%%

\section{Conclusions and discussion}
\label{conclusions}

In conclusion, we have performed microscopic calculations of
total and ionization energies in order to assess
the accuracy of the Faddeev RPA approach for light atoms.  

The FRPA method is an expansion of the many-body self-energy
that makes explicit the coupling between particles and collective phonons.
This formalism includes all contributions from perturbation theory up
to third order, which is crucial to a correct prediction of IEs for outer-valence
electrons in atomic and molecular systems.
At the same time, it also includes full resummations of RPA diagrams
necessary to screen the long-range Coulomb interaction 
in extended systems~\cite{Mattuk,DicVan}.
While the FRPA includes completely the ADC(3) theory, it is not
{\em a-priori} guaranteed that it performs equally well in small
systems since problems related to violations of
the Pauli principle---inherent with the RPA method---could worsen
in this situation.
Nevertheless, the present results show that this is
not the case and the approach is capable of good accuracy even for the
two-electron problem. %, He  and Be$^{2+}$.

 In general, it is found that  FTDAc/ADC(3)c  and  FRPAc give very similar 
results for the lightest systems while the inclusion of RPA theory for
phonons leads to small but systematic improvements as the atomic number
increases.
For total binding energies, their difference is negligible
in the He and Be atoms while the FRPAc yelds $\approx$5~mH
more correlation energy for atomic numbers Z$\geq$10.
Except for the lightest atoms, 99\% of the the total correlation energy
is normally recovered and the total energies obtained agree well with
CCSD (as expected~\cite{Tro.05}).
The discrepancies with the experimental data are also within the errors
estimated for the extrapolation to the basis set limit.
 The only notable exception is the neutral Be atom, for which the small gap at
the Fermi surface complicates the extraction of the correlation energy.
In this case, the discrepancy obtained with respect to the experiment
appears to be mostly due to deficiencies in the basis set. However,
a smaller fraction of it is probably related to missing correlations
and/or to the lack of full self-consistency in the FRPAc (HF reference
states were used).

Similar trends are found for the ionization energies. For
the two-electron cases, He and Be$^{2+}$, FRPAc does not introduce
improvements with respect to FTDAc/ADC(3)c but still gives sensible
predictions. The above problems in describing the correlations
of neutral Be are also reflected in the results for the first ionization
orbit.
 For all other cases, the introduction of RPA phonons shifts IEs
by 2-10~mH and always brings them closer to the experiment. On average,
the {\em rms} error for outer valence IEs lowers from 13.7 to 10.6~mH.
The $3s$ orbit in Ar is found to be fragmented and configuration mixing
effects between 2h1p states are required to obtain the correct ionizaion
energy and relative intensity.
%For the deeper $2p$ orbit in Ar a similar improvement from RPA is found,
%although the agreement  with the experiment worsens due to increasing
%importance of 3h2p configuration (neglected by both FDTA and FRPA).

Numerically, the FRPA can be implemented as a diagonalization in
2p1h-2h1p space, implying about the same cost as an ADC(3) calculation.
The present study was based on numerical codes originally developed
in nuclear physics and for spherical systems~\cite{Bar.01,DiB.04},
which are not suitable
for chemistry applications. Testing the FRPA approach on molecules
requires developing appropriate codes that handle molecular geometries. 
Work along this line is in progress and will be reported it in a
forthcoming publication~\cite{Deg.tbp}.

Due to the inclusion of RPA excitations, the FRPA method holds promise
of bridging the gap between accurate descriptions of quasiparticles
in finite and extended systems.
 Investigating the feasibility of FRPA for larger molecules and
the electron gas is therefore a priority for future research efforts.
Consistent calculations of quasiparticle properties in these cases,
once feasible, will be important for constraining
functionals in quasiparticle density functional theory~\cite{Van.06}.

\acknowledgments
 We acknowledge several useful discussions with Prof. W. H. Dickhoff.
 This work was supported by the Japanese Ministry of Education, Science
and Technology (MEXT) under KAKENHI grant no. 21740213. MD acknowledges support from FWO-Flanders. DVN and MD are members of the QCMM Alliance Ghent-Brussels.

\end{document}